\documentclass[prd,preprint, superscriptaddress,preprintnumbers,nofootinbib]{revtex4}
\usepackage{graphicx}
\usepackage{epsfig}
\usepackage{bm}
\usepackage{latexsym,amssymb,amsmath,amssymb,wasysym,float}

\usepackage{color}

\usepackage{enumitem}

\newcommand{\postscript}[2]{\setlength{\epsfxsize}{#2\hsize}
   \centerline{\epsfbox{#1}}}

\usepackage[usenames,dvipsnames]{xcolor}
\definecolor{orange}{cmyk}{0,0.5,1,0}
\definecolor{rossoCP3}{cmyk}{0,.88,.77,.40}
\definecolor{graa}{rgb}{0.8,0.8,0.8}
\definecolor{blaa}{rgb}{0.2,0.2,0.6}

\begin{document}

\preprint{MPP-2024-139}
\preprint{LMU-ASC 09/24}

\title{\color{rossoCP3} Bulk Black Hole  Dark Matter}

\author{Luis A. Anchordoqui}

\affiliation{Department of Physics and Astronomy,\\  Lehman College, City University of
  New York, NY 10468, USA
}

\affiliation{Department of Physics,\\
 Graduate Center,  City University of
  New York,  NY 10016, USA
}

\affiliation{Department of Astrophysics,
 American Museum of Natural History, NY
 10024, USA
}

\author{Ignatios Antoniadis}

\affiliation{High Energy Physics Research Unit, Faculty of Science, Chulalongkorn University, Bangkok 1030, Thailand}

\affiliation{Laboratoire de Physique Th\'eorique et Hautes \'Energies
  - LPTHE \\
Sorbonne Universit\'e, CNRS, 4 Place Jussieu, 75005 Paris, France
}

\author{Dieter\nolinebreak~L\"ust}

\affiliation{Max--Planck--Institut f\"ur Physik,  
 Werner--Heisenberg--Institut,
80805 M\"unchen, Germany
}

\affiliation{Arnold Sommerfeld Center for Theoretical Physics, 
Ludwig-Maximilians-Universit\"at M\"unchen,
80333 M\"unchen, Germany
}

\author{Karem Pe\~nal\'o Castillo}
\affiliation{Department of Physics and Astronomy,\\  Lehman College, City University of
  New York, NY 10468, USA
}

\begin{abstract}
\vskip 2mm \noindent The dark dimension provides a  mechanism to resolve the cosmological hierarchy problem and assembles a colosseum for dark matter contenders. In a series of recent publications we investigated whether primordial black
 holes (PBHs) perceiving the dark dimension could constitute all of
 the dark matter in the universe. A key assumption of these
 investigations is that PBHs remain confined to the brane during the
 entire evaporation process. As a consequence, the abundance of PBHs evaporating at the present epoch is severely constrained by observations of both the extragalactic and Galactic
$\gamma$-ray backgrounds. As a natural outgrowth of these
investigations, herein we relax the assumption of brane localized PBHs
and reexamine the evaporation process of PBHs which are allowed to
escape into the dark dimension. We show that the escape of PBHs from
the brane is almost instantaneous. Armed with this pivotal finding we
reassess the allowed mass range of PBHs to assemble all cosmological
dark
matter, which is estimated to be $10^{11}
\alt M_{\rm BH}/{\rm g} \alt 10^{21}$. 
\end{abstract}

\maketitle

\section{Introduction}

The set of effective field theories (EFTs) that look consistent
according to all available low-energy criteria, but do not arise from
an ultraviolet complete theory that includes quantum gravity has become known as
the swampland~\cite{Vafa:2005ui}. There are several conjectures for
fencing off the
swampland~\cite{Palti:2019pca,vanBeest:2021lhn,Agmon:2022thq,Anchordoqui:2024rov}. For
example, the distance conjecture (DC) asserts that in any consistent
theory of quantum gravity, when venturing to large distances in
four-dimensional (4D) Planck units within scalar field space, a tower of particles will become light at a rate that is exponential in the field space distance~\cite{Ooguri:2006in}. Concatenated to the DC is the anti-de
Sitter (AdS) distance conjecture, which
hitches the dark energy density to the mass scale $m$ characterizing the infinite tower of states,
$m \sim |\Lambda|^\alpha$, as the negative AdS vacuum energy
 $\Lambda \to 0$, with $\alpha$ a positive constant of ${\cal O}
 (1)$~\cite{Lust:2019zwm}. Furthermore, if this scaling behavior stays
 around valid in de Sitter (or
 quasi de Sitter) space, an unbounded number of massless modes also would pop
 up in the limit $\Lambda \to 0$.
 
The AdS-DC in dS
space provides an expressway to elucidate the cosmological hierarchy
problem, $\Lambda/M_p^{4} \sim 10^{-122}$, because it connects the
size of the compact space $R_\perp$ to the
dark energy scale $\Lambda^{-1/4}$ via
\begin{equation}
  R_\perp \sim \lambda \ \Lambda^{-1/4} \,,
\label{RperpLambda}  
\end{equation}  
where the proportionality factor is estimated to be within the range
$10^{-1} < \lambda < 10^{-4}$ and where $M_p$
is the Planck mass~\cite{Montero:2022prj}. As a matter of fact,
(\ref{RperpLambda}) stems from
  constraints of both theory and experiment. Firstly, the accompanying Kaluza-Klein (KK) 
  tower contains massive spin-2 bosons, and thus the Higuchi
  bound~\cite{Higuchi:1986py} sets an absolute upper bound to
  $\alpha$. Secondly, explicit string calculations
of the vacuum energy~(see
e.g.~\cite{Itoyama:1986ei,Itoyama:1987rc,Antoniadis:1991kh,Bonnefoy:2018tcp})
lead to a lower limit on $\alpha$. More precisely, these two stringy 
constraints imply $1/4 \leq \alpha \leq 1/2$. Now,  when these
theoretical arguments are combined with
experimental observations (e.g. constraints on deviations from Newton's
gravitational inverse-square law~\cite{Lee:2020zjt} and neutron star
heating~\cite{Hannestad:2003yd}) we arrive at the conclusion encapsulated
in (\ref{RperpLambda}): {\it The cosmological hierarchy problem can be
  addressed if there exists one extra dimension of
radius $R_\perp$ in the micron range, and the lower bound for $\alpha =
1/4$ is basically saturated~\cite{Montero:2022prj}.} A theoretical
amendment on the connection between the cosmological and KK mass scales confirms $\alpha =
1/4$~\cite{Anchordoqui:2023laz}. A second conclusion encapsulated in (\ref{RperpLambda}) is that the KK tower of the new (dark)
dimension lays bare at the mass scale $m_{\rm KK} \sim
1/R_\perp$.

The 5-dimensional Planck scale
(or species scale where gravity becomes
strong~\cite{Dvali:2007hz,Dvali:2007wp,Cribiori:2022nke,vandeHeisteeg:2023dlw,Dvali:2009ks,Dvali:2010vm, Dvali:2012uq, vandeHeisteeg:2022btw, Calderon-Infante:2023uhz,Castellano:2023aum,Cribiori:2023sch,Cribiori:2023ffn,Castellano:2023stg,Basile:2023blg,Basile:2024dqq,Bedroya:2024ubj,Herraez:2024kux})
 provides an upper bound for the ultraviolet cutoff of EFTs of gravity
 coupled to a number of light particle species and within the dark
 dimension scenario is found to be
\begin{equation}
M_* \sim m_{\rm KK}^{1/3} \
M_p^{2/3}\, .
\end{equation}

The dark dimension heralds a new era of particle phenomenology and
cosmology~\cite{Anchordoqui:2022ejw,Anchordoqui:2022txe,Blumenhagen:2022zzw,Anchordoqui:2022tgp,Gonzalo:2022jac,Anchordoqui:2022svl,Anchordoqui:2023oqm,vandeHeisteeg:2023uxj,
  Noble:2023mfw,Anchordoqui:2023wkm,Anchordoqui:2023tln,Anchordoqui:2023qxv,Law-Smith:2023czn,Anchordoqui:2023etp,Obied:2023clp,Anchordoqui:2023woo,Anchordoqui:2024akj,Vafa:2024fpx,Bedroya:2024uva,Anchordoqui:2024dxu,Gendler:2024gdo,Anchordoqui:2024gfa,Antoniadis:2023sya}. For
example, primordial black holes (PBHs) with a Schwarzschild radius smaller than a micron could be potential
dark matter candidates~\cite{Anchordoqui:2022txe,Anchordoqui:2022tgp,
  Anchordoqui:2024akj,Anchordoqui:2024dxu}. If such PBHs were
evaporating at the present epoch, then their abundance would be strongly
constrained by observations of both the extragalactic and Galactic
$\gamma$-ray backgrounds~\cite{Carr:2020xqk,Villanueva-Domingo:2021spv}. The null results from searches in $\gamma$-ray
experiments, imply that only for the  black hole mass range
\begin{equation}
  10^{15} \alt M_{\rm BH}/{\rm g} \alt 10^{21} \, ,
\label{MBHrange}
\end{equation}
PBHs could make all cosmological dark matter~\cite{Anchordoqui:2024dxu}. This
bound was derived within the probe-brane approximation, which ensures
that the black hole is bind to the brane
during the evaporation process.  However, a thorough study of the
evolution of the PBH-brane system, once the black hole is
given an initial velocity (that mimics the recoil effect due to
graviton emission in the bulk), suggests that the brane would bend
around the PBH forcing it to eventually escape
into the dark dimension once the two portions of the brane come in
contact and reconnect~\cite{Frolov:2002as,Frolov:2002gf,Flachi:2005hi,Flachi:2006hw}. Motivated by this pivotal finding,
in this paper we explore the possibility of enriching the dark
dimension with bulk black hole dark matter. 

The layout is as follows. We begin in Sec.~\ref{sec:2} with a synopsis
of PBH formation in standard 4D inflation and after that we lay out
the hypotheses for PBH formation within the dark dimension scenario. In Sec.~\ref{sec:3} we reexamine
the Hawking evaporation process considering 
particle emission onto the brane and in the bulk. In Sec.~\ref{sec:4}
we estimate the time scale for PBHs escaping the brane in the mass range
of interest. Armed with our finding, we reassess limits on the abundance of PBH dark
matter by focusing on black holes which could escape from the brane into the
bulk, and therefore would avoid constraints from $\gamma$-ray experiments. The
paper wraps up in Sec.~\ref{sec:5} with some conclusions.

Before proceeding, we pause to note that the dark dimension can be
understood as a line interval with end-of-the-world 9-branes in
M-theory attached
at each end, which is equivalent to a semicircular 11th dimension endowed
with $S^1/\mathbb{Z}_2$ symmetry~\cite{Schwarz:2024tet}. It is worthy of
mention that if the dark
dimension is warped then the brane could behave as if the tension were infinite, resulting in the impossibility for PBHs to leave the brane~\cite{Flachi:2006ev}.

\section{Primordial black hole production}

\label{sec:2}

Arguably, the most well-motivated model for PBH
production makes use of quantum fluctuations generated during
inflation~\cite{Zeldovich:1967lct,Hawking:1971ei,Carr:1974nx}. {\it Planck} measurements of the CMB tightly
constrain temperature fluctuations on large scales, because the power
spectrum of the comoving curvature perturbation ${\cal R}$ is almost scale invariant: ${\cal} P_{\cal R} (k) = {\cal A} (k/k_0)^{n_s-1}$, with 
amplitude $\ln(10^{10} {\cal A}) = 3.044 \pm 0.014$, spectral index $n_s =
0.965 \pm 0.004$,
and pivot scale $k_0 = 0.05~{\rm Mpc}^{-1}$~\cite{Planck:2018vyg}. However, observational constraints on smaller
scales (which cannot be probed by CMB data) are loose enough to allow for
large fluctuations, with amplitude of power spectrum ${\cal
  O}(0.01)$~\cite{Green:2020jor}. In this section we review the
necessary conditions for obtaining large curvature
fluctuations that could ignite 
PBH formation with appreciable abundance today.

To lay out the foundation of the discussion we first note that the
order of magnitude estimate of the black hole mass $M_{\rm BH}$ can be
guessed by equating the scaling of the cosmological energy density with time $t$
in the radiation dominated epoch,
\begin{equation}
  \rho \sim M_p^2/t^2 \,,
\label{rhouno}
\end{equation}
to the required density in a region of mass $M_{\rm BH}$ which is able to collapse within its Schwarzschild radius
\begin{equation}
\rho \sim  M_p^6/M_{\rm BH}^2 \, .
\label{rhodos}
\end{equation}
Drawing a connection between (\ref{rhouno}) and (\ref{rhodos}) we can
see that at production PBHs would
have roughly the cosmological horizon mass~\cite{Carr:2020xqk}
\begin{equation}
M_{\rm BH} \sim  t M_p^2 \sim 10^{15}
\left(\frac{t}{10^{-23}~{\rm s}}\right)~{\rm g} \, .
\label{yieldrho}
\end{equation}
Substituting in (\ref{yieldrho}) representative cosmological time scales we can roughly estimate that black
holes produced at the Planck time 
($10^{-43}~{\rm s}$) would have the Planck mass ($M_p
\sim 10^{-5}~{\rm g}$), black holes produced at the QCD epoch
($10^{-5}~{\rm s}$) would have a solar mass ($1~M_\odot$), and black
holes produced at $t \sim
1~{\rm s}$ would have a mass of $10^{5} M_\odot$, which is comparable
to the mass of the black holes thought to reside in galactic nuclei. The back-of-a-napkin calculation delivering (\ref{yieldrho}) suggests that PBHs could span an enormous mass
range. Even though the spectrum of masses of these PBHs remains
unspecified, on cosmological scales PBHs could behave like a typical cold dark
matter particle.

Taking into account the cosmological expansion, the
present day PBH density parameter in units of the critical density
$\Omega_{\rm PBH}$ can be related to the initial collapse fraction at redshift
$z$ by
\begin{equation}
 \beta = \frac{\Omega_{\rm PBH}}{\Omega_r} \ \frac{1}{(1+ z)} \sim
 10^{-6} \ \Omega_{\rm PBH} \ \left(\frac{t}{{\rm s}}\right)^{1/2} \sim
10^{-18} \ \Omega_{\rm PBH} \ \left(\frac{M_{\rm BH}}{10^{15}~{\rm g}}
  \right)^{1/2} \,,
\label{beta1}
\end{equation}  
where $\Omega_r \sim 10^{-4}$ is the present-day density parameter of
radiation and in the last step we have used
(\ref{yieldrho})~\cite{Carr:2005zd}. Note that the $(1 + z)$ factor derives from the fact that the radiation density scales
as $(1 + z)^4$, while the PBH density scales as $(1 + z)^3$. This
implies that $\beta (M_{\rm BH})$ must be tiny even if PBHs provide all of the dark matter~\cite{Carr:1975qj}.

The criterion for PBH formation is most easily specified in terms of
the density contrast $\delta \equiv (\rho - \bar
\rho)/\bar\rho$, where $\rho$ is the density of the region and $\bar
\rho$ is the average background density. A PBH will form if the
density contrast exceeds a critical value, $\delta_c$. The PBH abundance at formation epoch can then be interpreted as the
fraction of the universe with 
regions dense enough to produce PBHs, 
\begin{equation}
\beta \equiv  \int_{\delta_{\rm c}}^{\infty} P(\delta) \ d \delta,
\label{beta2}
\end{equation}
where $P(\delta)$ is the probability distribution function, which
describes the likelihood that a given fluctuation have an over-density
$\delta$, and we assume that a perturbation will collapse to form PBH
if its amplitude is larger than a critical value
$\delta_c$~\cite{Press:1973iz}. It is generally assumed that the density
fluctuations have a Gaussian distribution 
\begin{equation}
P_{G}(\delta)= \frac{1}{\sqrt{2\pi}\sigma} e^{-{(\delta -\mu)^2}/{2
    \sigma^2}},
\label{gaussian}
\end{equation}
and are spherically symmetric, where $\mu$ is the mean and $\sigma^2$
is the variance of the
distribution.  Equating (\ref{beta1}) to (\ref{beta2}), we can 
estimate the required mass variance $\sigma^2$ of (\ref{gaussian}) that can
give rise to large population of PBH today. Setting $\mu = 0$ and
integrating (\ref{beta2}) it follows that
\begin{equation}
\beta = \int_{\delta_{\rm c}}^{\infty} \frac{d
  \delta}{\sqrt{2 \pi} \sigma} \exp \left(-\frac{\delta^{2}}{2
    \sigma^{2}}\right) = \frac{1}{2} \ {\rm
  Erfc}\left(\frac{\delta_{\rm c}}{\sqrt{2}\sigma}\right)\simeq \frac{
  \sigma}{\sqrt{2 \pi} \delta_{\rm c} } \ \exp \left(-\frac{\delta_{\rm c} ^{2}}{2 \sigma^2}\right),
\end{equation}
where ${\rm Erfc}(x) = 1 - {\rm Erf}(x)$ is the complementary error
function and in the last equality we have assumed $\delta_{\rm c} \gg
\sigma$,  which is a good approximation for all practical
purposes. Indeed, following~\cite{Harada:2013epa} we set $\delta_c \sim 0.4$, and so for
an all-dark-matter interpretation
with $M_{\rm BH} \sim 10^{15}~{\rm g}$ we
arrive at $\sigma \sim 0.045$, where we have taken the present-day
dark-matter density from Planck's measurements $\Omega_{\rm DM} h^2 \sim
0.120 \pm 0.001$~\cite{Planck:2018vyg}, with a dimensionless Hubble constant $h \sim 0.73$~\cite{Riess:2021jrx,Murakami:2023xuy}. 

Now, we can estimate the primordial power spectrum of the curvature
perturbation at the time of PBH formation, which is of order the square of the mass variance,
\begin{equation}
  {\cal P}_{\cal R} (k_{\rm PBH}) \sim \sigma^2 \,,
\end{equation}
and so for Gaussian fluctuations the amplitude of the scalar power
spectrum needed at scales $k_{\rm PBH}$ relevant for PBH formation is
found to be ${\cal P}_{\cal R} (k_{\rm PBH}) \sim
10^{-3}$~\cite{Green:2024bam,Ozsoy:2023ryl}. We can then infer that the assumption $\Omega_{\rm PBH}
= \Omega_{\rm CDM}$ requires a very large amplification of the curvature spectrum between large CMB and small PBH-formation
scales; namely, 
\begin{equation}
  \Delta {\cal P}_{\cal R} \equiv \frac{{\cal P}_{\cal R} (k_{\rm PBH})}
  {{\cal P}_{\cal R} (k_{\rm CMB})} \sim 10^6 \, .
\end{equation}

Noticeably, one- and two-field slow-roll inflationary dynamics could actually generate a peak in the curvature power
spectrum~\cite{Ivanov:1994pa,Garcia-Bellido:1996mdl,Randall:1995dj,Garcia-Bellido:2017mdw}. This is possible if the effective dynamics of the inflaton
field presents a near-inflection point which slows down the field right before the end of inflation.
This simple and quite generic feature of the potential makes the
inflaton enter into an ultra-slow-rolling stage during a short range
of e-folds that gives rise to a prominent spike in the fluctuation power spectrum at scales much smaller than those
probed by CMB experiments. Concerns were raised regarding one-loop corrections to
the large power spectrum in single-field inflation~\cite{Kristiano:2022maq,Kristiano:2023scm}. However, these corrections have been found
to be negligible when the transition from the ultra-slow-roll to the
slow-roll phase is
smooth~\cite{Riotto:2023hoz,Riotto:2023gpm,Firouzjahi:2023aum,Firouzjahi:2023ahg}.\footnote{For
  additional input on the one-loop correction from the short modes
  onto the large-scale power spectrum,
  see~\cite{Choudhury:2023vuj,Motohashi:2023syh,Firouzjahi:2024psd,Choudhury:2024jlz,Ballesteros:2024zdp,Kristiano:2024ngc,Kristiano:2024vst}.}

An alternative proposal for PBH production within slow-roll inflation,
advocates for nucleation of bubbles of metastable vacuum during a time
interval that is short compared to
the
inflationary Hubble time~\cite{Kleban:2023ugf}. These bubbles
would then expand exponentially during inflation to super-horizon
size, and later collapse into black holes when the expansion of the
universe is decelerating. As a consequence, even though $M_{\rm BH}$ is exponentially sensitive to the moment bubbles form during inflation, the resulting
 PBH mass spectrum can be nearly monochromatic. If bubble nucleation 
 occurs near the middle of
 inflation, $M_{\rm BH}$ can fall in the range given in
 (\ref{MBHrange}), in which 
PBHs could make all cosmological dark matter.

The preceding discussion has been framed within the context of
standard 4D inflation. This implies that the size of the dark
dimension during the inflationary epoch should have been much smaller
than a micron (most likely at $M_*^{-1}$); otherwise, as we observed
in~\cite{Anchordoqui:2022svl}, the dark dimension scenario would be inconsistent with the Higuchi bound~\cite{Higuchi:1986py}. On the other
hand, to have 5D black hole formation and evaporation, one should
also assume that the size of the dark dimension became much bigger by
some 4D potential that was created immediately after the end of
inflation. Alternatively, it is always possible that the PBHs are
produced in the bulk to start with. This situation will be more
appealing within the proposal introduced elsewhere~\cite{Anchordoqui:2022svl,Anchordoqui:2023etp,Antoniadis:2023sya}, in which we
postulated that the dark dimension
may have undergone a uniform rapid expansion, together with the
three-dimensional non-compact space, by regular exponential inflation
driven by an (approximate) higher dimensional cosmological
constant.\footnote{Note that the amplitude of the power spectrum observed in the branes should be suppressed by the bulk volume in order to convert $M_*$ to 4D $M_p$.} If this were the case, then primordial fluctuations during inflation of the compact space could lead to the
production of black holes in the bulk. An investigation along these
lines is obviously important to be done, but it is beyond the scope of
this paper. Herein, we proceed under the hypothesis that 5D PBHs could
be formed within the dark dimension scenario. Shortly after the PBHs are formed they experience Hawking
evaporation. It is this that we now turn to study.

\section{Hawking Evaporation}
\label{sec:3}

In the mid-70's Hawking pointed out that a black hole emits thermal
radiation as if it were a black
body, with a temperature inversely
proportional to its mass~\cite{Hawking:1974rv,Hawking:1975vcx}. However, in the neighborhood of the horizon
the black hole produces an effective potential barrier that
backscatters part of the emitted radiation, modifying the thermal
spectrum. The so-called ``greybody factor'' $\sigma^{(s)} (\omega)$, which controls
the black hole absorption cross section, depends upon the spin of the
emitted particles $s$, their energy $\omega$, and $M_{\rm
  BH}$~\cite{Page:1976df,Page:1976ki,Page:1977um}. In this section we
carry out a numerical analysis of the evaporation of black holes 
perceiving the dark dimension to determine the emission rate onto the
brane and in the bulk.

We assume that the black hole is spherically symmetric and can be treated as a flat
$(4+n)$ dimensional object. This
assumption is valid for extra dimensions that are larger than the
$(4+n)$-dimensional Schwarzschild radius \begin{equation}
    r_s = \frac{1}{M_*} \left[ \frac{M}{M_*} \right]^{1/(n+1)}
    \left[ \frac{8 \, \Gamma\big(\frac{n+3}{2} \ \big)}{(n+2) \ \pi^{(n+1)/2} }
    \right]^{1/(n+1)} \,, 
\label{r_s}
  \end{equation}
where $\Gamma(x)$ is the Gamma
function~\cite{Tangherlini:1963bw, Myers:1986un,Argyres:1998qn}. The 
higher-dimensional Schwarzschild black hole behaves like a thermodynamic system~\cite{Hawking:1976de}, with
temperature $T_H \sim (n+1)/(4 \pi r_s)$ and entropy $S = (4 \pi
M_{\rm BH} r_s)/(n+2)$~\cite{Anchordoqui:2001cg}.

The emission rate of particles of
spin $s$ is found to be
\begin{equation}
\frac{dN^{(s)}}{dt} =  \sigma^{(s)}(\omega) \ \left[
\exp \left( \frac{\omega}{T_H} \right) - (-1)^{2s} \right]^{-1} 
\ \frac{d^{d-1}k}{(2\pi)^{d-1}}\,,
\end{equation}
where $k$ is the $(d-1)$-momenta of the particle living in $d$ dimensions, and
\begin{eqnarray}
\sigma^{(s)} (\omega) & = & \sum_l  \frac{2^{n+1} \ \pi^{(n+1)/2}\
                            \Gamma\left(\frac{n+3}{2}\right)}
{\omega^{n+2} \ n!}\ \frac{(2l+n+1) \ (l+n)!}{ l!}\,|{\cal
  A}^{(s)}_l (\omega)  |^2\nonumber \\
  & = &  \frac{2^{n}}{\pi} \
 \frac{\left\{\Gamma\left(\frac{n+3}{2}\right)\right\}^2 }{(\omega
        r_s)^{n+2}} \ \ A_{\rm
        hor} \  {\cal N}_l \
      |{\cal A}^{(s)}_l (\omega)|^2\,,
\end{eqnarray}      
and where
\begin{equation}
{\cal N}_l = \frac{(2l+n+1)\,(l+n)!}{l! \,(n+1)!}\,,
\end{equation}
 is the multiplicity of states corresponding to the
 same partial wave $l$,
\begin{eqnarray}
A_{\rm hor} &=& 
r_s^{n+2}\,\int_0^{2 \pi} \,d \varphi \,\prod_{k=1}^{n+1}\,
\int_0^\pi\,\sin^k\theta_{k+1}\,
\,d\theta_{k+1} = 2 \pi \ r_s^{n+2} \ \prod_{k=1}^{n+1} \ \sqrt{\pi}\,\,\frac{\Gamma(\frac{k+1}{2})}
{\Gamma(\frac{k+2}{2})}
                \nonumber \\
            &=& 2 \pi \ r_s^{n+2} \
\frac{\pi^{(n+1)/2}}{\Gamma\left(\frac{n+3}{2}\right)} 
\end{eqnarray}
is the horizon area of the $(4+n)$-dimensional black hole, and ${\cal A}^{(s)}_l (\omega)$ the absorption coefficient~\cite{Han:2002yy}. Note that the number of
dimensions $d$ in which a particular field lives should not be
confused with dimensionality of spacetime $(4 +n)$. For massless particles, $|k|=\omega$, and the phase-space
integral reduces to an integral over the energy of the emitted
particle $\omega$. For massive particles, $|k|^2=\omega^2 -m^2$, which
implies that for a black hole to emit a particle of mass $m$, its
temperature $T_H \geq m$. 

Bearing this in mind, the evaporation rate of a given 
particle species of spin $s$ in the frequency window $(\omega, \omega+d\omega)$ is given by~\cite{Anchordoqui:2002cp}
\begin{equation}
\frac{d^2 N^{(s)}}{dt \, d\omega} = \frac{1}{2\pi}
\sum_{l} {\cal N}_l \ |{\cal A}^{(s)}_l (\omega)|^2\ \left[
\exp \left( \frac{\omega}{T_H} \right) - (-1)^{2s} \right]^{-1}
\, .
\label{dNdtdomega}
\end{equation}
The absorption coefficients for different particle species have been
calculated in~\cite{Kanti:2002nr,Kanti:2002ge} and for emission of
particles on the brane, in the low frequency limit $\omega r_s \ll 1$, are given by
\begin{equation}\label{sca_brane}
|{\cal A}_l^{(0)} (\omega)|^2= \frac{16 \pi}{(n+1)^2}
\left(\frac{\omega r_s}{2}\right)^{2 l+2}
\frac{\left\{\Gamma(\frac{l+1}{n+1}) \right\}^2 \  \left\{\Gamma(1+\frac{l}{n+1}) \right\}^2}{\left\{\Gamma(\frac{1}{2}+l) \right\}^2 \ \left\{\Gamma(1 +\frac{2l+1}{n+1})\right\}^2} \,,
\end{equation}
\begin{equation}\label{ferm_brane}
|{\cal A}^{(\frac{1}{2})}_{l} (\omega)|^2 = \frac{2\pi \ (\omega
  r_s)^{2l+1} \ 2^{-(4l+2)/(n+1)}}
{2^{2l} \ \left\{\Gamma(l+1) \right\}^2} \,,
\end{equation}
and 
\begin{equation}
|A^{(1)}_{l} (\omega)|^2\ = \frac{(2\omega r_s)^{2+ 2l}}{(n+1)^2}\,
\frac{\left\{\Gamma\left(\frac{l}{n+1}\right)\right\}^2 \ 
\left\{\Gamma\left(\frac{l+1}{n+1}\right)\right\}^2\
\left\{\Gamma(l+2)\right\}^2}
{\left\{\Gamma\left(\frac{2l+1}{n+1}\right)\right\}^2 \ \left\{\Gamma(2l+2) \right\}^2} \, .
\end{equation}
The absorption coefficient for graviton emission in the bulk is
estimated to be
\begin{equation}
|{\cal A}^{(s)}_{l} (\omega)|^2=  4\pi \left(\frac{\omega r_H}{2}\right)^{2l+n+2}
\frac{\left\{\Gamma\left(1+\frac{l}{n+1}-G\right)\right\}^2\ \left\{\Gamma\left(1+\frac{l}{n+1}+G\right)\right\}^2}
{\left\{\Gamma\left(l+\frac{n+3}{2}\right)\right\}^2 \
  \left\{\Gamma\left(1+\frac{2l}{n+1}\right)\right\}^2}\, ,
\label{abs3}
\end{equation}
where
\begin{equation}
    G =\frac{1}{2(n+1)}\,\sqrt{(n+2)^2 - \frac{q + \tilde p + \tilde w}
{[2m + (n+2) (n+3)]^2}}
\label{G_S}
\end{equation}
for $s=0$, and 
\begin{equation}
    G=\frac{(1+k)(n+2)}{4(n+1)}\,
\label{G_TV}
\end{equation}
for $s=1$ and $s=2$, with 
$m \equiv l(l+n+1) - n -2$,
 $q \equiv (n+2)^4 (n+3)^2$,  
$z \equiv 16 m^3 + 4m^2 (n+2)(n+4)$,
$p \equiv (n+2) (n+3) \left[4m\,(2n^2+5n+6) +n (n+2) (n+3) (n-2)\right]$,
$w \equiv -12m\,(n+2)\,\left[m (n-2) + n (n+2) (n+3)\right]$,
   $\tilde p = p - z (n+2)^2 (n+3)^2/(4 m^2)$, and 
$\tilde w = w - z (n+2) (n+3)/m$~\cite{Creek:2006ia}.

\begin{figure}
    \centering
    \includegraphics[width=\textwidth]{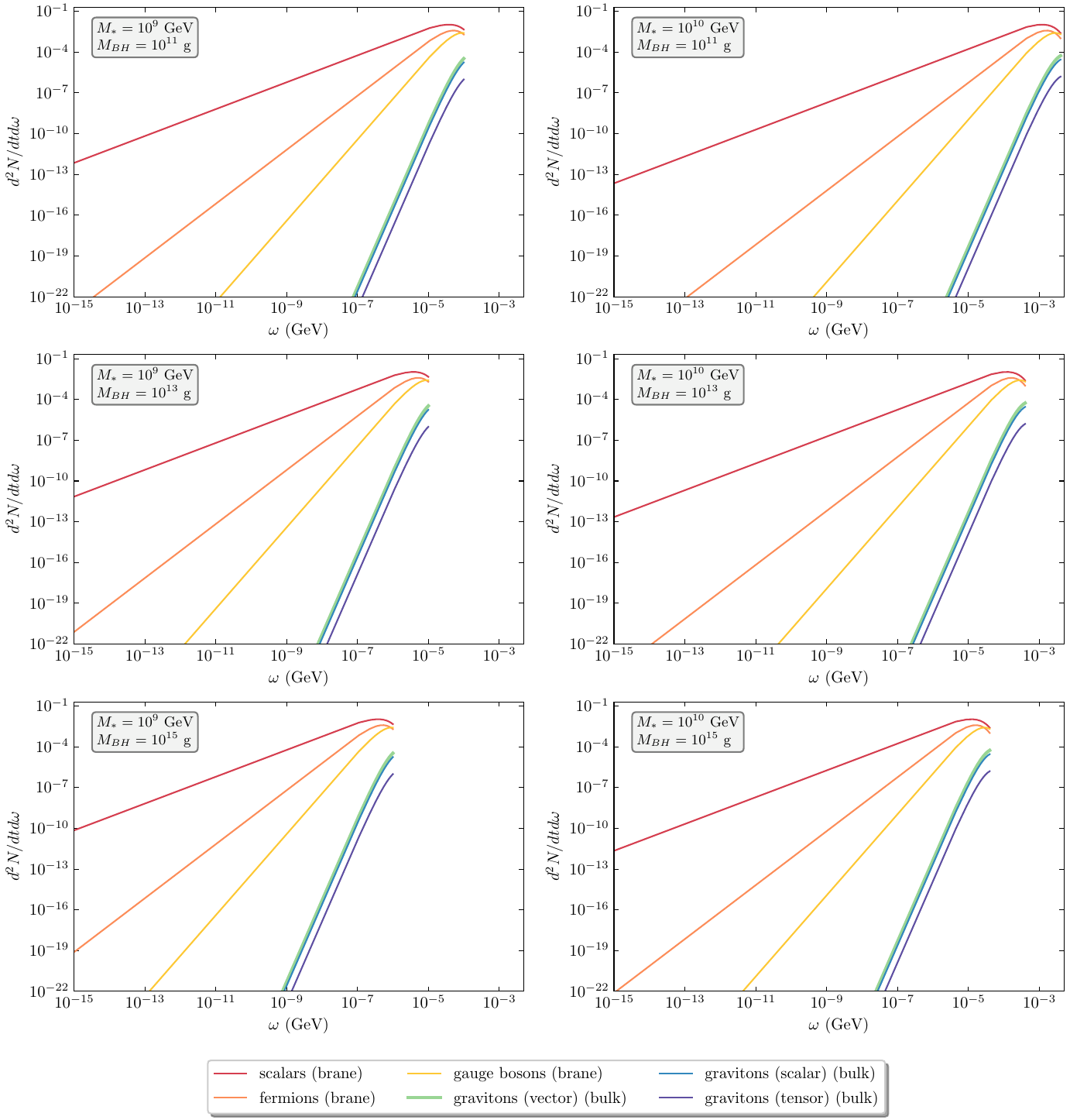}
    \caption{Evaporation rate for 
particles of spin $s$. We have taken $n=1$, a species scale $M_* = 10^9~{\rm GeV}$ (left) and
$M_* = 10^{10}~{\rm GeV}$ (right), considering from top to bottom
$M_{\rm BH}/{\rm g} = 10^{11},\, 10^{13},\, 10^{15}$.}
    \label{fig:1}
\end{figure}

In Fig.~\ref{fig:1} we show the evaporation rate for particles of spin
$s$ from PBHs perceiving the dark dimension, considering benchmark
black hole masses. One can check by inspection that for large values
of $\omega$, the differences in emission rate for all brane localized
species become smaller, as one would expect because in the high
frequency limit, the greybody factors for all species approach the
geometric optics limit. The curves have been truncated at the
frequency in which the low frequency approximation, adopted to derive 
the greybody factors, breaks down. Note
that the
radiation of particles with higher spin is suppressed at low frequencies because of the
larger barrier such particles have to surmount. Note also that $d^2N^{(s)}/dt\, d\omega$ depends
only weakly on $M_{\rm BH}$ and $M_*$. For a comparison with previous
estimates of the particle emission rate, see Appendix.

\section{Filling the Dark Matter Void with Bulk Black Holes}
\label{sec:4}

In this section, in line with our stated plan, we turn to revise previous
estimates of the fraction of dark matter that could be composed of PBHs. Higher-dimensional black holes radiate
Standard Model fields and gravitons on the brane as well as gravitons into the
bulk. The total power emitted in both channels scales as
\begin{equation}
 - \frac{dM_{\rm BH}}{dt} \sim \kappa_n \ T_H^{n+4} \ r_s^{n+2} \sim
 \kappa_n \ T_H^2 \,,
\label{power}
\end{equation}
where $\kappa_n$ encodes the effect of greybody factors~\cite{Ireland:2023zrd}. As can be seen in Fig.~\ref{fig:1}, $\kappa_n$ depends only weakly on $T_H$.

A point worth noting at this juncture is that during the process of
black hole evaporation $M_{\rm BH}$ decreases and $T_H$ rises. In what follows, we adopt a quasi-stationary approach 
to the evaporation process, which implies that the black hole has time to come into 
equilibrium at each new temperature before the next particle is
emitted. We also rely on the semi-classical assumption of
self-similarity; namely, we assume that in the course of
evaporation, a black hole gradually shrinks in size while maintaining
the standard semi-classical relations between its parameters, such as
its mass, the Schwarzschild radius, and the Hawking temperature. Note
that this is a very strong assumption, because we will estimate the
evaporation rate exclusively for a black hole of a fixed radius. With
this assumption in mind, (\ref{power})
can be solved for the lifetime
\begin{equation}
\tau_{\rm BH} \sim \frac{1}{\kappa_n \ M_*} \ \left(\frac{M_{\rm
      BH}}{M_*}\right)^{(n+3)/(n+1)} \, .
\label{lifetime}
\end{equation}

Now, the recoil effect due to graviton emission imparts the black hole
a relative kick velocity $v$ with respect to the brane, allowing the PBH
to escape into the bulk. The time scale of the escape of a black hole
is estimated to be~\cite{Flachi:2005hi}
\begin{equation}
  \tau_{\rm esc} \sim \frac{1}{\kappa_n} \ \frac{r_s}{v} \, .
\label{escape}
\end{equation}  
If the origin of the recoil is the inhomogeneous emission of particles
during black hole evaporation, a rough estimate of the kick velocity
gives 
\begin{equation}
  v \sim \frac{f_{M_{\rm BH}}}{ \sqrt{N_{\rm emitted}}} \ ,
  \label{kickv}
\end{equation}
where $f_{M_{\rm BH}}$ is the fraction of the evaporated mass and
$N_{\rm emitted} \sim f_{M_{\rm BH}} M_{\rm BH}/T_H$ is the number of
emitted particles~\cite{Kovacik:2021qms}. Substituting (\ref{r_s}) and (\ref{kickv}) into (\ref{escape}) it follows that
\begin{equation}
  \tau_{\rm esc} \sim 
  \frac{1}{\kappa_n \ \sqrt{f_{M_{\rm BH}}} M_*} \left(\frac{M_{\rm
        BH}}{M_*} \right)^{(n+4)/(2n+2)} \, .
\end{equation}
Now, $f_{M_{\rm BH}}$ at $\tau_{\rm esc}$ is estimated to be
$f_{M_{\rm BH}} \sim \tau_{\rm esc}/\tau_{\rm BH}$, and so we arrive at
\begin{equation}
  \tau_{\rm esc} \sim \frac{1}{ \kappa_n  \ M_*} \ \left(\frac{M_{\rm
        BH}}{M_*} \right)^{(2n+7)/(3n+3)} 
\end{equation}
and
\begin{equation}
  f_{M_{\rm BH}} \sim \left(\frac{M_*}{M_{\rm BH}}
  \right)^{(n+2)/(3n+3)} \, .
\end{equation}
Taking $n=1$ and $M_* \sim 10^{10}~{\rm GeV}$, we can conclude that PBH
of $M_{\rm BH} \sim 10^{15}~{\rm g}$ would almost immediately escape
the brane.

To a very good approximation we can then assume the black holes will
only emit gravitons in the bulk, and so the semi-analytic studies carried out in~\cite{Friedlander:2022ttk,Friedlander:2023qmc}
 yield $\kappa_1 \sim 0.06$.\footnote{This 
   semi-analytic estimate is in agreement with those derived
   in~\cite{Harris:2003eg,Cardoso:2005mh}.} Now, setting $n=1$, $M_*
 \sim 10^{10}~{\rm GeV}$, and considering the age of the universe of 13.8~Gyr~\cite{Planck:2018vyg} we can parametrize (\ref{lifetime}) as
\begin{equation}
  \tau_{\rm BH} \sim 13.8 \ \left(\frac{\kappa_1}{0.06}\right)^{-1}  \ \left(\frac{M_{\rm
        BH}}{10^{11.37}~{\rm g}}\right)^2~{\rm Gyr} \, .
\end{equation}
Because PBHs evaporate almost entirely into gravitons in the bulk, the constraints
from gamma-ray
observations~\cite{Carr:2020xqk,Villanueva-Domingo:2021spv} can be
safely neglected. Taking into account the systematic error associated with the order of magnitude
uncertainty in the species scale $10^9 \alt M_*/{\rm GeV} \alt 10^{10}$, we
conclude that the mass
range for PBHs to make all the cosmological dark matter is then 
\begin{equation}      
 10^{11}  \alt M_{\rm BH}/{\rm g} \alt 10^{21} \, .
\end{equation}  
Note that the minimum $M_{\rm BH}$ required for bulk PBHs to make all
cosmological dark matter is in agreement with
the estimate derived in~\cite{Anchordoqui:2024dxu} using scaling arguments.

In closing, we stress that if there were 5D primordial
near-extremal black holes in nature, then it would be possible to
lower the minimum mass allowing a PBH all-dark-matter interpretation,
because near-extremal black holes are colder and
longer-lived~\cite{Anchordoqui:2024akj}.
Note, however, that near-extremal black holes cannot escape from the
brane into the
bulk since they carry electromagnetic charge. It is also important to
stress 
that if PBHs experience the memory burden effect~\cite{Dvali:2018xpy,Dvali:2020wft}, then the minimum
$M_{\rm BH}$ allowing a 5D PBH all-dark-matter interpretation can also be
relaxed~\cite{Alexandre:2024nuo,Dvali:2024hsb,Thoss:2024hsr,Balaji:2024hpu}. Note that the limits from $\gamma$-ray observations adopted
in~\cite{Anchordoqui:2024dxu} to constrain this possibility would be
evaded if PBHs escape
from the brane.   

\section{Conclusions}
\label{sec:5}

We have reexamined the Hawking evaporation process of PBH perceiving
the dark dimension. We have shown that PBHs are expected to escape from
the brane almost instantaneously. Motivated by this essential finding,
we have revised the allowed mass range of PBHs that could assemble all cosmological
dark matter. We found that a PBH all-dark-matter interpretation would be possible in the mass range  $10^{11}
\alt M_{\rm BH}/{\rm g} \alt 10^{21}$.

We end with an observation. Complementary to the PBHs, it was observed in~\cite{Gonzalo:2022jac} that the universal
coupling of Standard Model fields to the massive spin-2 KK excitations of the graviton in the
dark dimension provides an alternative  dark matter contender. The
cosmic evolution of the dark graviton gas is primarily dominated by
``dark-to-dark'' decays, yielding a specific realization of the
dynamical dark matter framework~\cite{Dienes:2011ja}. A remarkable close
relation between PBHs and the dark gravitons has been pointed out
in~\cite{Anchordoqui:2022tgp}. We note that the ideas discussed in
this paper make more noticeable
the PBH $\leftrightharpoons$ dark graviton gas connection, as in both
scenarios the
cosmological evolution of the dark
matter sector would take place mostly in the bulk. Actually, PBHs
originally formed on the brane could act as a source of KK
gravitons. Note that for $M_{\rm BH}$ in the range of interest the
Hawking temperature,
\begin{equation}
  T_H \sim \frac{1}{r_s}  \sim \left(\frac{M_{\rm BH}}{10^{12}~{\rm
          g}}\right)^{-1/2}~{\rm MeV} \,,
\end{equation}
is consistent with experimental constraints on the natural scale in
which the KK gravitons on the dark dimension are generated with enough
abundance~\cite{Gonzalo:2022jac,Law-Smith:2023czn,Obied:2023clp}.

\section*{Acknowledgements}
The work of L.A.A. and K.P.C. is supported by the U.S. National Science
Foundation (NSF Grant PHY-2112527). I.A. is supported by the Second
Century Fund (C2F), Chulalongkorn University.  The work of D.L. is supported by the Origins
Excellence Cluster and by the German-Israel-Project (DIP) on Holography and the Swampland.

\section*{Appendix}

\begin{figure}[htb!]
  \begin{minipage}[t]{0.49\textwidth}
    \postscript{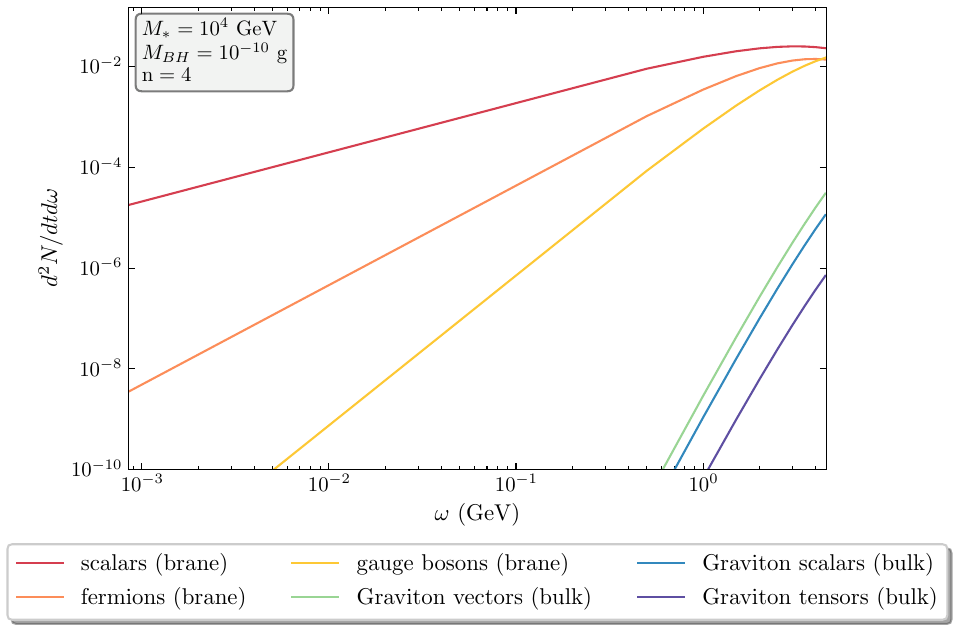}{0.99}
  \end{minipage}
\begin{minipage}[t]{0.49\textwidth}
    \postscript{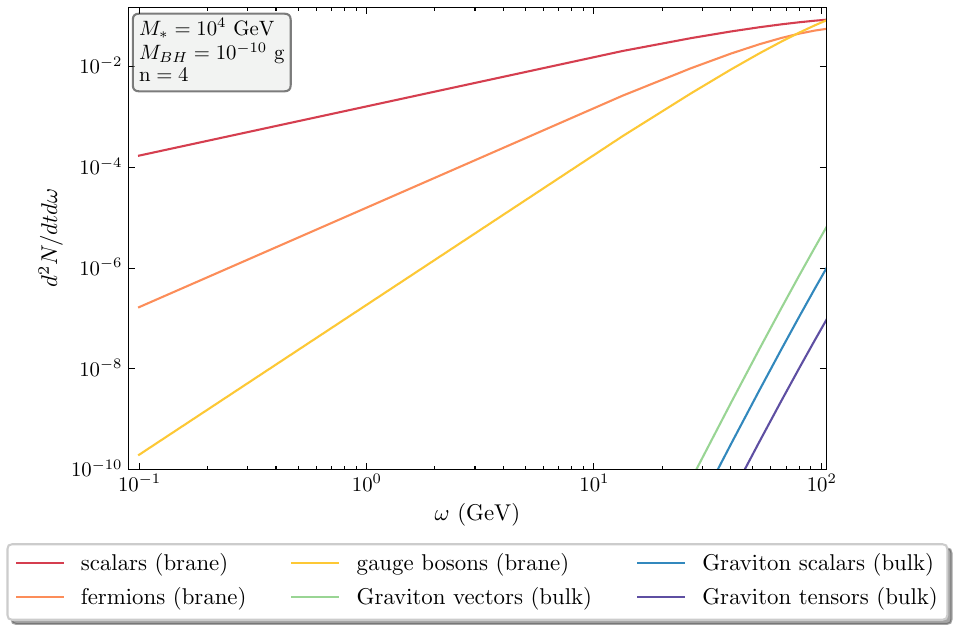}{0.99}
  \end{minipage}
  \caption{Particle emission rate of
    Schwarzschild, near extremal, and quantum
    black holes; for $d=4$ (left) and $d=5$ (right). \label{fig:2}}
  \end{figure}

In this Appendix we compare our estimates of
$d^2N^{(s)}/dt d\omega$ with those obtained in~\cite{Ireland:2023zrd} for
$n=2$ and $n=4$. In Fig.~\ref{fig:2} we show the instantaneous emission of
species on the brane and in the bulk for two and four extra dimensions, with
$M_{\rm BH} = 10^{-10}~{\rm
  g}$ and $M_* = 10~{\rm TeV}$. One can check by inspection of Fig.~\ref{fig:2}
and Fig.~1 in~\cite{Ireland:2023zrd} that there is good agreement
between the two estimates.

\end{document}